\documentclass[aps,nofootinbib,prd,eqsecnum,showpacs,showkeys,preprintnumbers]{revtex4-2}
\usepackage{graphicx}
\usepackage{wrapfig}
\usepackage{caption}
\usepackage{subcaption}
\usepackage{amsmath}
\usepackage{amsfonts}
\usepackage{amssymb}
\usepackage{color}
\usepackage{bm}
\usepackage{natbib}
\usepackage{epstopdf}
\usepackage{url}
\usepackage{footnote}
\usepackage{textcomp}
\usepackage{ulem}
\usepackage{esint}
\usepackage[unicode=true, pdfusetitle,
 bookmarks=true,bookmarksnumbered=false,bookmarksopen=false,
 breaklinks=false,pdfborder={0 0 1},backref=false,colorlinks=false]{hyperref}
\usepackage{multirow}
\usepackage{pifont}
\usepackage{lipsum}
\usepackage[mathlines]{lineno}
\usepackage[newcommands]{ragged2e}
\begin{document}
\title{Primordial black holes within Higgs hybrid metric-Palatini approach}
\author{Brahim Asfour}
\email{brahim.asfour@ump.ac.ma}
\author{Farida Bargach}
\email{f.bargach@ump.ac.ma}
\author{Yahya Ladghami}
\email{yahyaladghami@gmail.com}
\author{Ahmed Errahmani}
\email{ahmederrahmani1@yahoo.fr}
\author{Taoufik Ouali}
\email{t.ouali@ump.ac.ma}
\affiliation{Laboratory of Physics of Matter and Radiation, \\
University of Mohammed first, BP 717, Oujda, Morocco\\
Astrophysical and Cosmological Center, Faculty of Sciences, BP 717, Oujda, Morocco} 

\date{\today }

\begin{abstract}
In this paper, we investigate the production of primordial black holes (PBHs) during the radiation-dominated era. The collapse of significant density perturbations originating from large primordial scalar fluctuations generated during inflation can lead to the formation of primordial black holes. In our study, we adopt the Higgs hybrid metric-Palatini model as our framework, in which the inflaton field and the Palatini curvature are non-minimally coupled. To achieve our objective, we analyze the behavior of the primordial curvature power spectrum, which exhibits a large enhancement at small scales corresponding to large wavenumbers $k$. Furthermore, we examine the probability of PBHs formation by studying the mass variance, $\sigma(M_{PBH})$, and the mass fraction of the total energy density collapsing into PBHs, $\beta(M_{PBH})$. The evolution of both functions is consistent with current observational constraints. Finally, we investigate the abundance of primordial black holes as a dark matter candidate. We found that they can account for the totality or a fraction of the current dark matter content, depending primarily on the values of the coupling constant and the e-folds number.   
\end{abstract}
%\pacs{98.80.Cq, 04.50.+h}
\maketitle
%%%%%%%%%%%%%%%%%%%%%%%
 
 \section{Introduction}
The concept of cosmic inflation has become the most widely accepted paradigm for studying the early Universe \cite{Guth,Starobinsky,Linde2,Albrecht,Sato,Linde1,Bargach:2019pst,Bouabdallaoui:2022wyp,Bouabdallaoui:2016izz}. It successfully addresses key cosmological issues, such as the horizon and the flatness problems. Moreover, it provides a natural mechanism for generating primordial curvature perturbations, which serve as the seeds of the large-scale structures observed in the present Universe. These perturbations originate from quantum fluctuations of the inflaton, the scalar field responsible for the early accelerated expansion.
\par On the other hand, the idea of the primordial black holes (PBHs) generation from the gravitational collapse of over-dense regions in the early Universe was first proposed by Zel'dovich and Novikov in 1967 \cite{Zeldovich:1967lct}. Subsequently, Hawking and Carr developed this concept through their influential research in the 1970s \cite{Hawking:1971ei,Carr:1974nx,Carr:1975qj}. In general, production of primordial black holes during the radiation-dominated era requires sufficiently large amplitude of primordial curvature perturbations generated during inflation. When these enhanced perturbations re-enter the horizon during the radiation era, the resulting over-dense regions can gravitationally collapse, leading to the formation of PBHs. Recently, PBHs have gained considerable attention as a potential source of gravitational waves (GWs), following the detection of GWs from the merger of two massive black holes (approximately 30 times the solar mass) by the Laser Interferometer Gravitational-Wave Observatory (LIGO) and Virgo collaborations \cite{LIGOScientific:2016aoc,LIGOScientific:2016sjg,LIGOScientific:2016lio,LIGOScientific:2017ycc,LIGOScientific:2018mvr}. Moreover, primordial black holes  are associated with a range of gravitational wave (GW) signals, including those generated by binary merger events as well as stochastic processes in the early Universe \cite{LISACosmologyWorkingGroup:2023njw}. In particular, stellar-mass PBHs have been proposed as potential sources of the GW events detected by LIGO and Virgo Collaborations. 
\par Additionally, the potential formation and properties of primordial black holes have been extensively studied within  the framework of various modified gravity theories. Among these theories are the Gauss-Bonnet (GB) inflation model \cite{Kawai:2021edk,Zhang:2021rqs}, the generalized Brans-Dicke theory \cite{Yi:2022anu} and the bouncing cosmology scenario \cite{Banerjee:2022xft}. Evidently, these investigations reveal that the single field inflationary models in the modified background dynamics can lead to a significant enhancement of primordial scalar perturbations at small scales, inducing a large curvature power spectrum. Consequently, PBHs can provide an additional probe to test and constrain modified gravity theories \cite{Papanikolaou:2021uhe,Papanikolaou:2022hkg,Lola:2020lvk,Geng:2015fla,Asfour:2025stv}.

\par The main purpose of this study is to investigate the production of primordial black holes during the radiation-dominated era, resulting from enhanced primordial curvature perturbations. To this end, we adopt the Higgs hybrid metric-Palatini inflation model, where the scalar field is non-minimally coupled to the Palatini term, as a framework. This hybrid approach is based on a combination between elements from both formalisms, metric and Palatini, to overcome the shortcomings of each pure theory \cite{Harko:2011nh, Capozziello:2015lza, He:2022xef, Shahid:2023dea}. In the metric formalism, all gravitational degrees of freedom are carried by the metric field and the connection is assumed to be Levi-Civita \cite{Bauer:2010jg,Rubio:2018ogq}. Conversely, the Palatini approach treats the metric and the connection as independent variables \cite{Tenkanen:2020dge}. This inflationary model and its applications have been discussed in detail in our previous works \cite{Asfour:2022qap,Asfour:2024mfr,Asfour:2025myu}. 

\par Meanwhile, the nature of dark matter (DM) remains one of the major open questions in modern cosmology. Although its existence has been inferred since 1930s following Fritz Zwicky's first observations of galaxy cluster dynamics, the fundamental nature of DM is still unknown \cite{Peebles:2017bzw}. In this work, we explore the possibility that DM is composed mainly of primordial black holes. Indeed, PBHs satisfy the key requirements for DM candidates: they are cold, non-baryonic, and stable. In particular, sufficiently massive PBHs could constitute either the entirety or a fraction of the current DM if their masses fall in the suitable region set by observations \cite{Chapline:1975ojl,Carr:2016drx,Carr:2020xqk,Ballesteros:2017fsr,Carr:2020gox,Belotsky:2014kca}. Notably, within the mass range of PBHs referred to as the "golden window" spanning between $10^{-16}$ and $10^{-11}$M$_s$ (with M$_s$ is the solar mass), primordial black holes could account for the totality of DM. Beyond this, PBHs may also contribute to the formation of large-scale structures through Poisson fluctuations \cite{Afshordi:2003zb} and act as seeds for supermassive black holes found in the galactic centers.

\par This work is organized as follows: in section \ref{secII}, we briefly review the basic equations that describe the Higgs hybrid metric-Palatini model. In section \ref{secIII}, we study the probability of PBHs production. Then, we investigate the possibility of PBHs to be a candidate of current DM in section \ref{secVI}. Finally, we discuss and conclude our results in section \ref{secV}.

\section{Higgs inflation dynamics} \label{secII}
We consider the hybrid metric-Palatini model in which the inflaton is non-minimally coupled to the Palatini curvature. The  action describing our model is given by \cite{ Asfour:2022qap}  
\begin{equation}
 S=\int d^4x \sqrt{-g}\left[ \frac{1}{2 \kappa^2}R + \frac{1}{2} \xi \phi^2 \hat{R}-\frac{1}{2} \partial_{\mu} \phi \partial^{\mu} \phi - V(\phi)\right], \label{eq2.1}
\end{equation}
where $R$ and $\hat{R}$ are the Einstein-Hilbert curvature and the Palatini curvature, respectively. The metric tensor $g_{\mu \nu}$ and the connection $\Gamma_{\beta \gamma}^{\alpha}$ are treated as independent variables. We define $\kappa^2=M_{Pl}^{-2}=8\pi G$ where $M_{Pl}$ denotes the reduced Planck mass, $\xi$ is the coupling constant and $V(\phi)$ represents the scalar field potential.
\par Considering the flat Friedmann-Robertson-Walker (FRW) background, Friedmann equations and Klein-Gordon equation are given respectively by \cite{ Asfour:2022qap}
\begin{equation}
H^2=\frac{\kappa^2}{3F(\phi)}\left[ \left( \frac{1}{2}-3\xi\sigma\right) {\dot{\phi}}^2+V(\phi)-6H\xi(1+\sigma) \phi \dot{\phi}\right], \label{eq2.3}
\end{equation}
\begin{equation}
 \square \phi  + \xi \hat{R}\phi - V_{\phi}=0,
\end{equation}
where $H=\dot{a}/a$ is the Hubble parameter, and $\sigma =0,1$ corresponds to the metric and Palatini cases, respectively. Here, we adopt $\sigma =1$. We also note that $ V_{\phi}=dV/d\phi$ is the derivative of the potential with respect to the scalar field $\phi$, and $F(\phi)$ is a function of scalar field translating the presence of the non-minimal coupling between the scalar field and the Palatini curvature, it is expressed as follows
\begin{equation}
F(\phi)=1+\xi\kappa^2 \phi^2.
\end{equation} 
Applying slow-roll conditions on Eq.\eqref{eq2.3}, we obtain the reduced form as  \cite{ Asfour:2022qap}
\begin{equation}
H^2	\simeq\frac{\kappa^2 V(\phi)}{3F(\phi)}, \label{eq2.5}
\end{equation}
and we express slow-roll parameters as  \cite{ Asfour:2022qap}
\begin{eqnarray}
 &\epsilon&=\frac{1}{2\kappa^2}\left( \frac{V_{\phi}}{V}\right)^2 Q, \label{eq2.6} \\
  &\eta&=\frac{ V _{\phi\phi}}{3H^2},\\
 &\zeta&=6\xi\sigma,\\
 &\chi&=\kappa^2_{eff}\left(  (1+2\xi-4\xi\sigma)\dot{\phi}-2\xi(1+\sigma)H\phi\right)
 \frac{(1+4\xi -2\xi\sigma)\dot{\phi}+10\xi(1+\sigma) H\phi}{2 F(\phi) H^2},
\end{eqnarray}
where $\kappa^2_{eff}=\kappa^2/\left[ 1+\frac{C\kappa^2}{2F(\phi)H}\phi \dot{\phi}\right]$ represents the effective gravitational constant and the term $Q$ indicates the correction to the standard case given as
\begin{equation}
Q=\frac{F(\phi)}{\alpha}\left( 1-\frac{4\xi\kappa^2\phi}{F(\phi)}\frac{V}{V_{\phi}}\right) \left( 1-\frac{2\xi\kappa^2\phi}{F(\phi)}\frac{V}{V_{\phi}}\right),
 \end{equation}
with $\alpha=1-6\xi\sigma$ and $C=2\xi(1+\omega)$.\\
\par Standard slow-roll parameters \cite{Liddle:1994dx} are recovered if we take $\xi=0$. The number of e-folds between horizon crossing and the end of inflation is defined as
\begin{equation}
 N=\int_{t_k}^{t_{end}}Hdt=\int_{\phi_k}^{\phi_{end}}\frac{H}{\dot{\phi}}d\phi, \label{eq2.11}
\end{equation}
where the indices "$k$" and "$end$" indicate the horizon crossing and the end of inflation, respectively.

\par The scalar spectral index of the power spectrum and the tensor-to-scalar ratio are expressed, respectively, as follows \cite{ Asfour:2022qap}
\begin{equation}
  n_s=1-2\epsilon+\frac{2}{\alpha}\left(\eta-\frac{\zeta}{3}-2\chi\right), \label{eq2.12}
\end{equation}
and
\begin{eqnarray}
r=\frac{1}{2\kappa^2}\frac{V^2_\phi}{ V^2} \frac{\left[ 1+\frac{C\kappa^2}{2F(\phi)H}\dot{\phi}\phi\right]^2}{(1-6\xi\omega)^2}. \label{eq2.13}
\end{eqnarray}

\subsection*{Higgs Potential} 
To check our finding, we consider the Higgs field as the inflaton. It is associated with the following quartic potential    
\begin{equation}
V(\phi)=\frac{1}{4}\lambda \phi^4, 
\end{equation}
where $\lambda$ is the Higgs self-coupling. Consequently, from Eqs. \eqref{eq2.12} and \eqref{eq2.13}, the scalar spectral index and the tensor-to-scalar take the form
\begin{eqnarray}
 n_s=&1&- \frac{16}{\alpha \kappa^2 \phi^2} (1 - \frac{\xi \kappa^2  \phi^2}{2 F(\phi)}) \\ 
\nonumber &+&\frac{2}{\alpha } \left[ \frac{12 F(\phi)}{\kappa^2  \phi^2} - 2 \xi \sigma - \kappa_{eff} ((1 + 2 \xi - 4 \xi \sigma) \dot{\phi} - 2 \xi (1 + \sigma) H \phi)\frac{ (1+4\xi -2\xi \sigma) \dot{\phi} +10 \xi(1+ \sigma) H \phi}{F(\phi) H}\right], \label{eq2.16}
\end{eqnarray}
and
\begin{equation}
r=\frac{8}{\kappa^2\phi^2}\frac{\left[ 1+\frac{ C\kappa^2}{2F(\phi)H}\dot{\phi}\phi\right]^2}{(1-6\xi\omega)^2},
\end{equation}
respectively. The number of e-folds can be expressed from Eq. \eqref{eq2.11} as
 \begin{eqnarray}
N_k=\frac{\alpha\kappa^2}{8}\left[ \phi_k^2-\phi_{end}^2\right]. \label{eq2.18}
\end{eqnarray}
\captionsetup[figure]{justification=Justified}
\begin{figure}[!htb]
\centering
\includegraphics[width=0.45\linewidth]{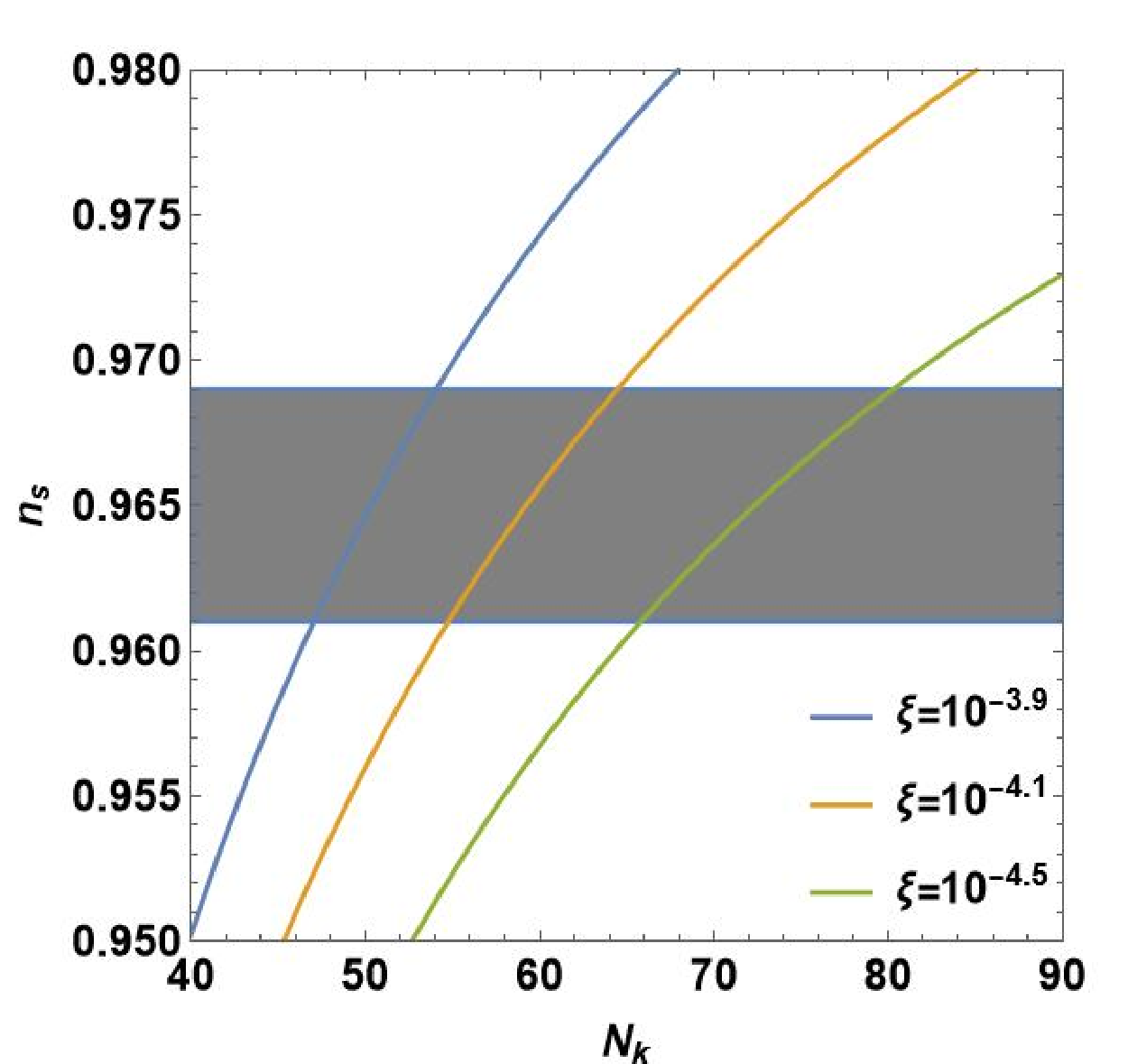} 
\caption{Variation of the scalar spectral index versus the e-folding number for three different values of the coupling constant $\xi=10^{-3.9}$, $\xi=10^{-4.1}$ and $\xi=10^{-4.5}$. The gray region indicates the Planck bounds imposed on $n_s$.}
\label{Fig:1}
\end{figure}
\begin{figure}[h!]
\centering
\begin{subfigure}[b]{0.49\textwidth}
\includegraphics[width=1\linewidth]{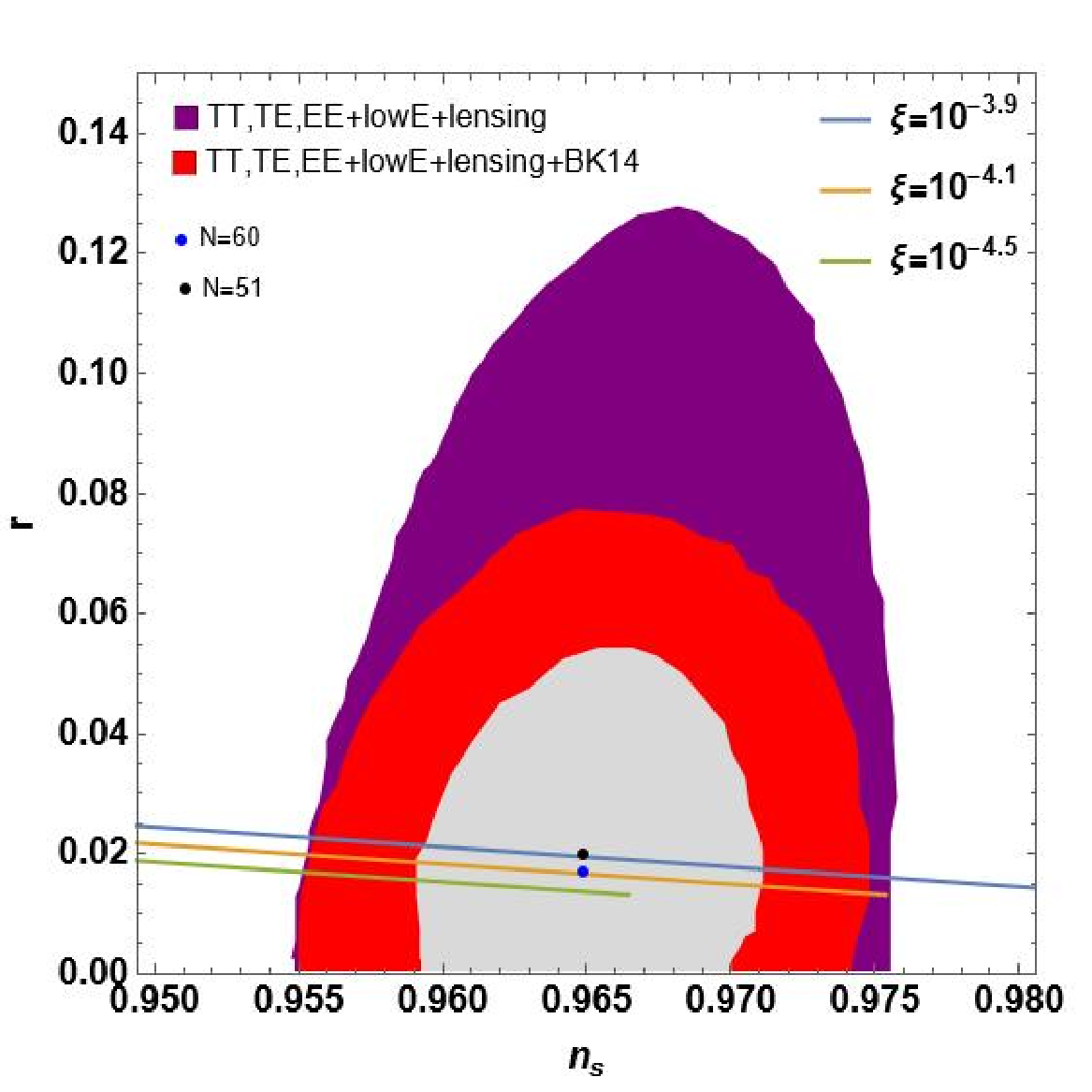}
\caption*{-2.a-}
\end{subfigure}
\begin{subfigure}[b]{0.49\textwidth}
\includegraphics[width=1\linewidth]{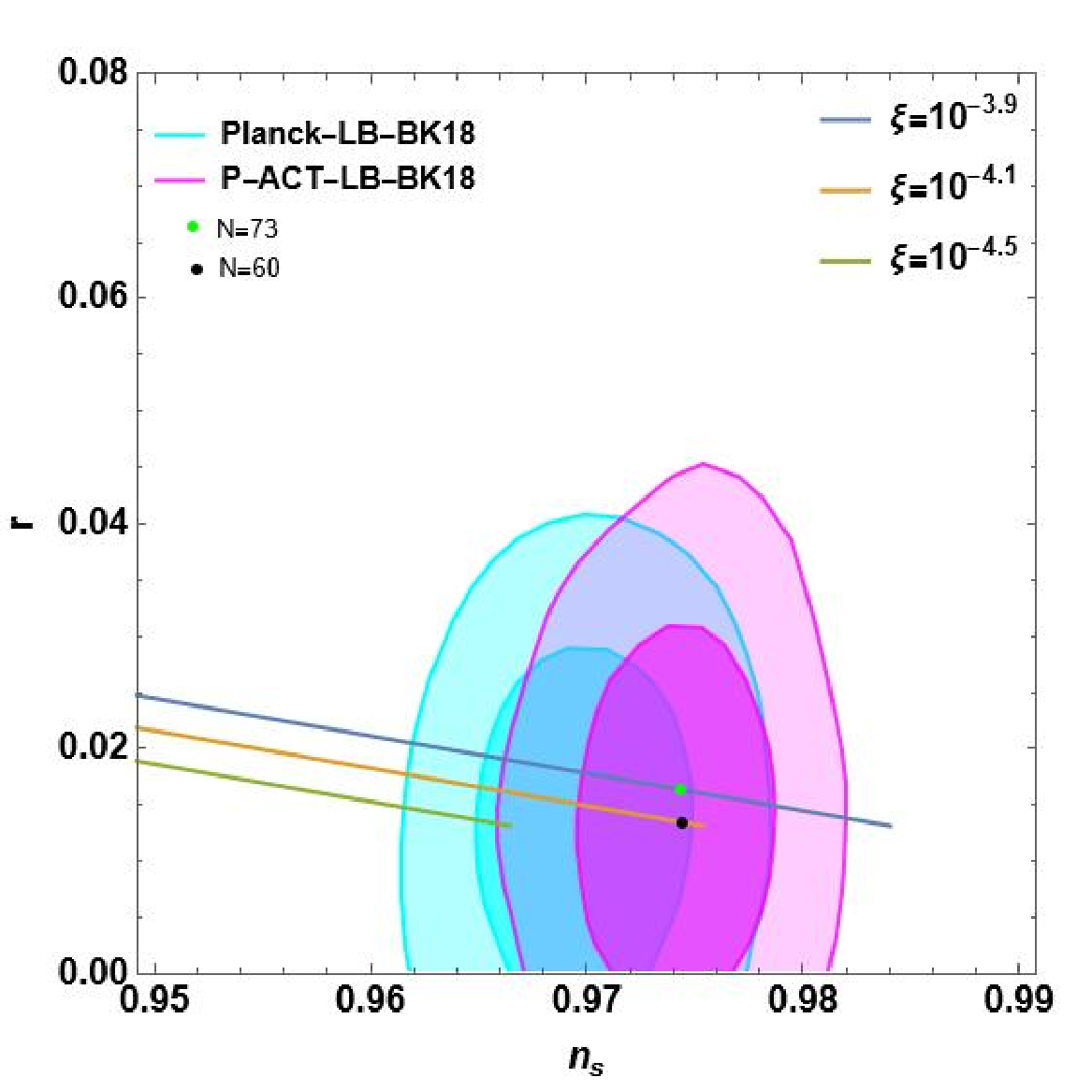}  
\caption*{-2.b-}
\end{subfigure} 
\caption{Evolution of the tensor-to-scalar, $r$, versus the spectral index, $n_s$, for different values of the coupling constant $\xi$, varying the e-folds number $N$ within the range $35\leq N\leq 75$.}
\label{Fig:2}
\end{figure}
Fig. $\ref{Fig:1}$ illustrates the variation of the spectral index, \( n_s \), versus the e-folds number considering different values of the coupling constant $\xi$ with $\lambda=10^{-9}$. The gray region indicates constraints imposed on the spectral index by the recent Planck data \cite{Planck:2018jri}. Our results are consistent with observational data as all curves fall within the bound for allowed range of number of e-folds. \par
In Fig. \(\ref{Fig:2}\), we show the $(n_s, r)$ plane for different values of the coupling constant $\xi$, varying the e-folds number within the range $35\leq N\leq 75$. The left panel includes observational constraints from the Planck TT,TE,EE+LowE+lensing (purple contour) and Planck TT,TE, EE+lowE+lensing+BK14 data (red contour) \cite{Planck:2018jri}. We find that theoretical predictions are compatible with the Planck data as all curves corresponding to the coupling constant values traverse the allowed regions. In the right panel, we confront the $(n_s, r)$ plane with the Planck (Planck-LB-BK18) and the combined Planck+ACT (P-ACT-LB-BK18) observational data. These constraints include BICEP/Keck CMB 2018 B-mode polarization measurements and are further enhanced by CMB lensing and by baryon acoustic oscillation data from DESI Year-1 \cite{AtacamaCosmologyTelescope:2025blo,AtacamaCosmologyTelescope:2025nti,DESI,DESI2}. It is evident that P-ACT-LB-BK18 dataset favors a slightly higher value of $n_s$ (with central value around $0.9743\pm0.0043$) and stronger upper limit on $r$ ($r<0.038$), than those preferred by Planck alone. As a result, both values of the coupling constant, $\xi=10^{-3.9}$ and $\xi=10^{-4.1}$, exhibit excellent consistency with the observations, as the corresponding curves comfortably cross the $1\sigma$ and $2\sigma$ confidence regions of both datasets with longer inflationary durations than those inferred from Planck alone.  In contrast, the case with $\xi=10^{-4.5}$ is strongly disfavored, as its curve lies almost entirely outside the 95\% confidence contours. Additionally, smaller values of $\xi$ lead to a smaller predicted tensor-to-scalar ratio $r$.

\section{Primordial black holes production}\label{secIII}
In the early radiation-dominated Universe, the collapse of large density perturbations, which is originated from large primordial scalar fluctuations produced during inflation, can lead to the formation of primordial black holes \cite{Zeldovich:1967lct,Hawking:1971ei,Carr:1975qj}. To investigate the possibility of primordial black hole production, it is necessary to compute the power spectrum of curvature perturbations and analyze its behavior on both small and large scales. The objective is to determine whether its evolution satisfies the conditions required for PBH formation. The primordial curvature power spectrum is defined as \cite{Papanikolaou:2022did}
\begin{eqnarray}
P_{R}(k)= \left(\frac{k^3}{2\pi^{2}}\right) {\lvert R \rvert}^2,
\end{eqnarray}
where $R$ is the comoving curvature perturbations given by \cite{ Asfour:2022qap}
\begin{eqnarray}
R=\Psi+\frac{H}{\dot{\phi} \left[ 1+\frac{C\kappa^2}{2F(\phi)H}\phi \dot{\phi}\right]} \delta\phi,
\label{eq2.19}
\end{eqnarray}
where $\Psi$ is a scalar perturbation. Considering the spatially flat gauge where $\Psi=0$, and using the relation between the scalar field fluctuations and the Mukhanov-Sasaki variable, $\delta\phi=v/a$, the comoving curvature perturbations $R$ can be written as
\begin{eqnarray}
R= \frac{H^2}{\dot{\phi}\sqrt{2k^3}}\left[ 1+\frac{C\kappa^2}{2F(\phi)H}\phi \dot{\phi}\right]^{-1} \left(\frac{k}{aH} \right)^{3/2-\nu}.
\end{eqnarray}
Here, the Mukhanov-Sasaki variable $v$ takes the form
\begin{eqnarray}
v=\frac{aH}{\sqrt{2k^3}}\left( \frac{k}{aH}\right)^{3/2-\nu}, \label{eq2.20}
\end{eqnarray}
where the parameter $\nu$ is expressed in terms of slow-roll parameters as
\begin{equation}
  \nu=\frac{3}{2}+\epsilon-\frac{1}{(1-6\xi\chi)}\left(\eta-\frac{\zeta}{3}-2\chi\right) .
\end{equation}
\begin{figure}[h!]
\centering
\begin{subfigure}[b]{0.49\textwidth}
\includegraphics[width=1\linewidth]{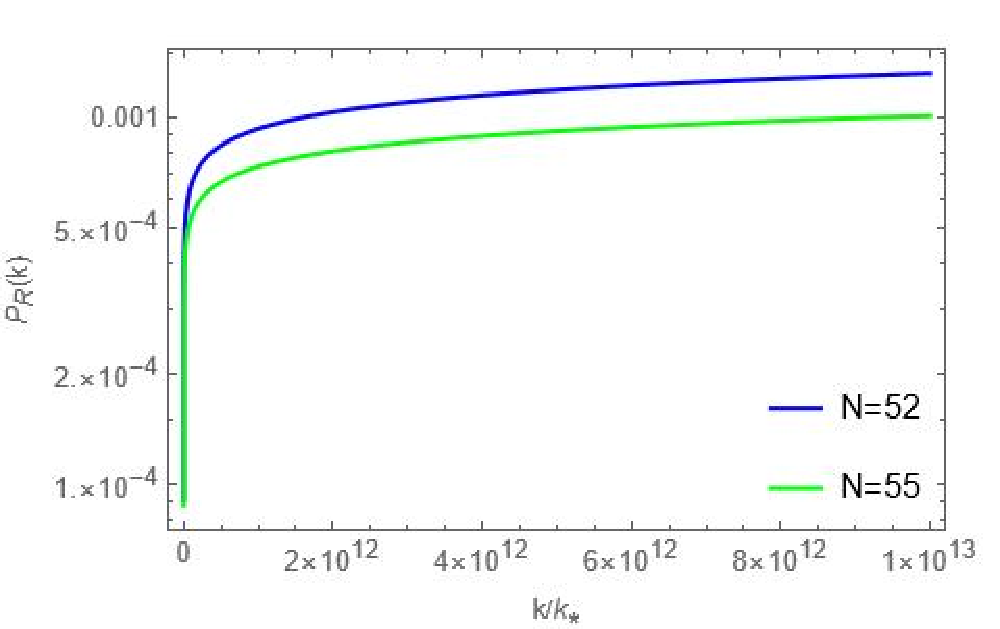}  
\caption*{-3.a-}
\end{subfigure}
\begin{subfigure}[b]{0.49\textwidth}
\includegraphics[width=1\linewidth]{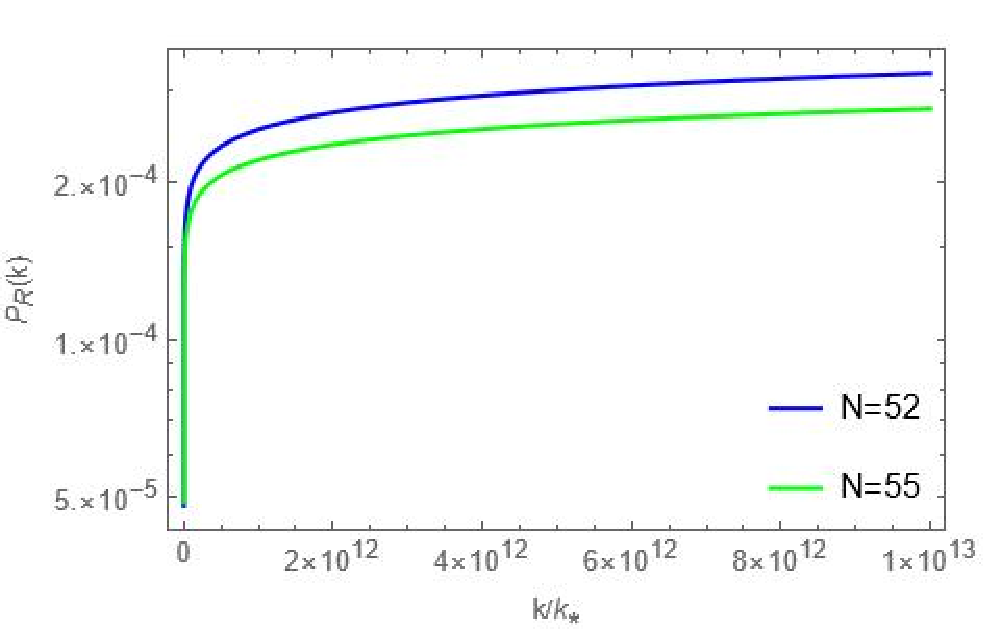} 
\caption*{-3.b-}
\end{subfigure} 
\caption{The primordial curvature power spectra, $P_{R}(k)$, as a function of the dimensionless ratio $k/k_{*}$. The blue  and the green lines correspond to the value of e-folding number, $N=52$ and $N=55$, respectively. We consider two values of the coupling constant, $\xi=10^{-4}$ (left panel) and $\xi=10^{-4.1}$ (right panel). Here, the pivot scale is taken as $k_{*}=0.05$ Mpc$^{-1}$.}
\label{Fig3}
\end{figure}
The evolution of the primordial curvature power spectrum, $P_{R}(k)$, is depicted in Fig. \ref{Fig3}, where it is plotted as a function of the dimensionless ratio, $k/k_{*}$. Two different values of the e-folding number, $N$, are considered: $N=52$ (blue line) and $N=55$ (green line). We also take into account two different values of the coupling constant, $\xi=10^{-4}$ and $\xi=10^{-4.1}$, presented in the left and the right panels, respectively. As evident, a significant enhancement in the curvature power spectrum at small scales with large wavenumber $k$ is observed. This implies the possibility of gravitational collapse of over-dense regions after inflation leading to the formation of the primordial black holes during radiation-dominated era. Furthermore, the enhancement of $P_{R}(k)$ is larger for smaller value of both $\xi$ and $N$.
\par The mass of the primordial black hole generated at time $t$ during the radiation-dominated era, $M_{PBH}$, is given by \cite{Sasaki:2018dmp,Mishra:2019pzq}
\begin{equation}
M_{PBH}(k)=\gamma M_{H}=4\pi \gamma \frac{M^2_{Pl}}{H}, \label{eq3.1}
\end{equation}
where $\gamma$ is the efficiency factor, taken to be $0.2$ \cite{Carr:1975qj}, represents the fraction of the mass within the Hubble horizon that collapses to form PBHs, $M_{H}$ is the horizon mass and $H$ is the Hubble parameter at the time of PBHs formation. Since the production of primordial black holes occurs during the radiation-dominated era, we can consider $\rho_{tot}\simeq\rho_{r}$ and hence write 
\begin{eqnarray}
H^2=\frac{\rho_{tot}}{3 M^2_{Pl}}=\frac{\rho_{r}}{3 M^2_{Pl}}=\Omega_{r,0}H^2_0\frac{\rho_r}{\rho_{r,0}}, \label{eq3.2}
\end{eqnarray}
where we have used the current energy density parameter of radiation
\begin{eqnarray}
\Omega_{r,0}=\frac{1}{3 M^2_{p}H^2_0}\rho_{r,0},
\end{eqnarray}
with $\rho_{r,0}$ and $H_0$ correspond to the present time radiation energy density and Hubble parameter, respectively. The energy density and temperature are related by the following two expressions
\begin{equation}
\rho_{r}=\frac{\pi^2}{30}g_* T^4,
\end{equation}
and
\begin{equation}
\rho_{r,0}=\frac{\pi^2}{30}g_{*,0} T^4_0,
\end{equation}
where $g_*$ and $g_{*,0}$ denote the effective numbers of relativistic degrees of freedom at the time of formation and at the present time, respectively. Thus we obtain
\begin{equation}
\frac{\rho_r}{\rho_{r,0}}=\frac{g_*}{g_{*,0}}\left( \frac{T}{T_0}\right)^4. \label{eq3.6}
\end{equation}
Under the assumption of entropy conservation, $dS=0$, implying that
\begin{eqnarray}
 S=sa^3=\text{constant},
\end{eqnarray}
with $s=\frac{2 \pi^2}{45} g_{*,s} T^3$, we get
\begin{eqnarray}
g_{s,*}T^3 a^3=g_{s,0}T^3_0 a^3_0,
\end{eqnarray}
where $g_{s,*}$ and $g_{s,0}$ refers to the effective numbers of relativistic degrees of freedom contributing to the entropy density at the time of formation and at the present-day, respectively. Note also that $a$ is the scale factor at the PBHs formation time and $a_0$ is the current scale factor. Then we can write the ratio of temperatures as 
\begin{eqnarray}
\left(\frac{T}{T_0}\right)^4=\left( \frac{g_{s,0}}{g_{s,*}}\right)^{4/3} \left(\frac{a_0}{a}\right)^{4}.
\end{eqnarray}
We substitute this expression into Eq.\eqref{eq3.6}, we get
\begin{eqnarray}
\frac{\rho_r}{\rho_{r,0}}=\frac{g_*}{g_{*,0}}\left( \frac{g_{s,0}}{g_{s,*}}\right)^{4/3}\left(\frac{a_0}{a}\right)^{4}.
\end{eqnarray}
By assuming $a_0$ to be unity and adopting the approximation $g_{s,*}=g_*$ during the radiation era, we return to Eq.\eqref{eq3.2} then we find
\begin{eqnarray}
H=H_0 \frac{\Omega_{r,0}^{1/2}}{a^{2}} \left(\frac{g_{*,0}}{g_{*}}\right)^{1/6}. \label{eq3.10}
\end{eqnarray}
Inserting this equation into Eq.\eqref{eq3.1}, we obtain
\begin{eqnarray}
M_{PBH}=\frac{4\pi \gamma M_{Pl}^2}{\Omega_{r,0}^{1/2}H_0}   \left( \frac{g_{*,0}}{g_{*}}\right)^{-1/6} a^2. \label{eq3.11}
\end{eqnarray}
The wavenumber associated  with horizon re-entry is given by the relation $k=a H$, using this relation, we can rewrite Eq.\eqref{eq3.10}
\begin{equation}
H=\frac{k^2}{\Omega_{r,0}^{1/2}H_0}\left(\frac{g_{*,0}}{g_{*}}\right)^{-1/6}. \label{eq3.12}
\end{equation}
Returning to Eq.\eqref{eq3.1}, one can express the mass of the generated primordial black hole, $M_{PBH}$, as a function of the corresponding comoving wavenumber $k$ as follows
\begin{eqnarray}
M_{PBH}(k)=4\pi \gamma M_{Pl}^2 \Omega_{r,0}^{1/2}H_0 \left( \frac{g_{*,0}}{g_{*}}\right)^{1/6} k^{-2}.
\end{eqnarray}
We simplify this expression by considering the following numerical values: $g_{*,0}=3.36$, $\Omega_{r,0}=9\times 10^{-5}$ and $H_0=100$ h km s$^{-1}$ Mpc$^{-1}$. Consequently, we get
\begin{equation}
M_{PBH}(k)\simeq M_{s} \left( \frac{\gamma}{0.2}\right) \left(\frac{106.75}{g_{*}}\right)^{1/6}\left( \frac{k}{1.6\times 10^6\text{Mpc}^{-1}}\right)^{-2}.  \label{eq3.14}
\end{equation}
Where the solar mass is $M_{s}=2\times 10^{33}$g, and the effective number of relativistic degrees of freedom is set to $g_{*}=106.75$.\par
Additionally, observational constraints imposed on the PBHs abundance indicate that their contribution is estimated to be less than $10^{-20}$ of the energy density of the Universe \cite{Sanchez:1997sv,Green:1997sz}. The fraction of the energy density that collapses into PBHs which measures the contribution of PBHs to the total energy density of the Universe is expressed using Press-Schechter formalism, as \cite{Green:2000he,Bullock:1998mi}              
\begin{eqnarray}
\beta\left( M_{PBH}\right)=\frac{\rho_{PBH}}{\rho_{tot}}=\int_{\delta_c}^{\infty} P(\delta)d\delta, \label{eq3.15}
\end{eqnarray}
where the probability distribution of the fluctuations, $P(\delta)$ is assumed to be Gaussian and given by \cite{Torres-Lomas:2013uzl}
\begin{eqnarray}
P(\delta) = \text{erfc}\left( \frac{\delta_c}{\sqrt{2}\, \sigma\left( M_{\text{PBH}} \right)} \right),
\end{eqnarray}
the complementary error function, \text{erfc} (....), can be approximated to yield to the following expression 
\begin{eqnarray}
P(\delta)=\frac{1}{\sqrt{2\pi}\, \sigma\left( M_{\text{PBH}} \right)} \exp\left( \frac{-\delta^2}{2\, \sigma^2\left( M_{\text{PBH}} \right)} \right),
\end{eqnarray}
where $\delta$ is the density contrast and $\delta_{c}$ indicates the critical value that the density fluctuation must exceed to undergo gravitational collapse and form a primordial black hole. Early studies have adopted $\delta_{c}=\omega$ (e.g. Ref. \cite{Carr:1975qj}). More recently, $\delta_{c}$ has been studied numerically in \cite{Niemeyer:1999ak,Musco:2004ak}. In this work, we consider $\delta_{c}$ as a function of the equation of state $\omega$  \cite{Harada:2013epa}
\begin{equation}
\delta_{c}=\frac{3(1+\omega)}{5+3\omega}\text{sin}^2\left( \frac{\pi\sqrt{\omega}}{1+3\omega}\right).
\end{equation}
Since the formation of primordial black holes happens in the radiation-dominated era, we take $\omega=1/3$ and hence $\delta_{c}\simeq 0.414$ \cite{Harada:2013epa}. The function $\sigma\left( M_{PBH}\right)$ represents the mass variance of the density perturbation defined by the primordial power spectrum curvature perturbations, $P_{R}(\tilde{k})$, and the window function $W^2(\tilde{k} /k)$ as \cite{Zhou:2020kkf}           
\begin{eqnarray}
\sigma^2\left( M_{PBH}\right)=\frac{4(1+\omega)^2}{(5+3\omega)^2} \int_{0}^{\infty} {\left(\frac{\tilde{k}}{k}\right) }^4 P_{R}(\tilde{k}) W^2(\tilde{k} /k) \frac{d\tilde{k}}{\tilde{k}}.
\end{eqnarray}
We take a Gaussian form for $W^2(\tilde{k} /k)$ as
\begin{eqnarray}
W(\tilde{k}/k)=\exp \left(\frac{-(\tilde{k}/k)^2}{2}\right).
\end{eqnarray}
In Press-Schechter theory \cite{Press:1973iz}, The mass variance can be expressed by the following relation as \cite{Bhattacharya:2021wnk}
\begin{eqnarray}
\sigma\left( M_{PBH}\right)\simeq \frac{2(1+\omega)}{3+5\omega}\sqrt{P_{R}(k)}.
\end{eqnarray}
According to observable constraints, $\sigma\left( M_{PBH}\right)$ must satisfy the condition $\sigma\left( M_{PBH}\right)<\sigma_{thresh}$, where $\sigma_{thresh}$ denotes the critical value above which the overproduction of PBHs would occur. In the non-Gaussian scenario, this threshold is set to $0.03$. However, within the Gaussian distribution approximation considered in this work, we have $\sigma_{thresh}=0.08$ \cite{Green:2000he}. This Gaussianity assumption for the curvature perturbations is widely adopted for its simplicity in studying PBHs production in single-field models. It remains reasonable, since interactions which could generate significant non-Gaussianity are suppressed by the slow-roll parameters. 

\par
\begin{figure}[h!]
\centering
\begin{subfigure}[b]{0.49\textwidth}
\includegraphics[width=1\linewidth]{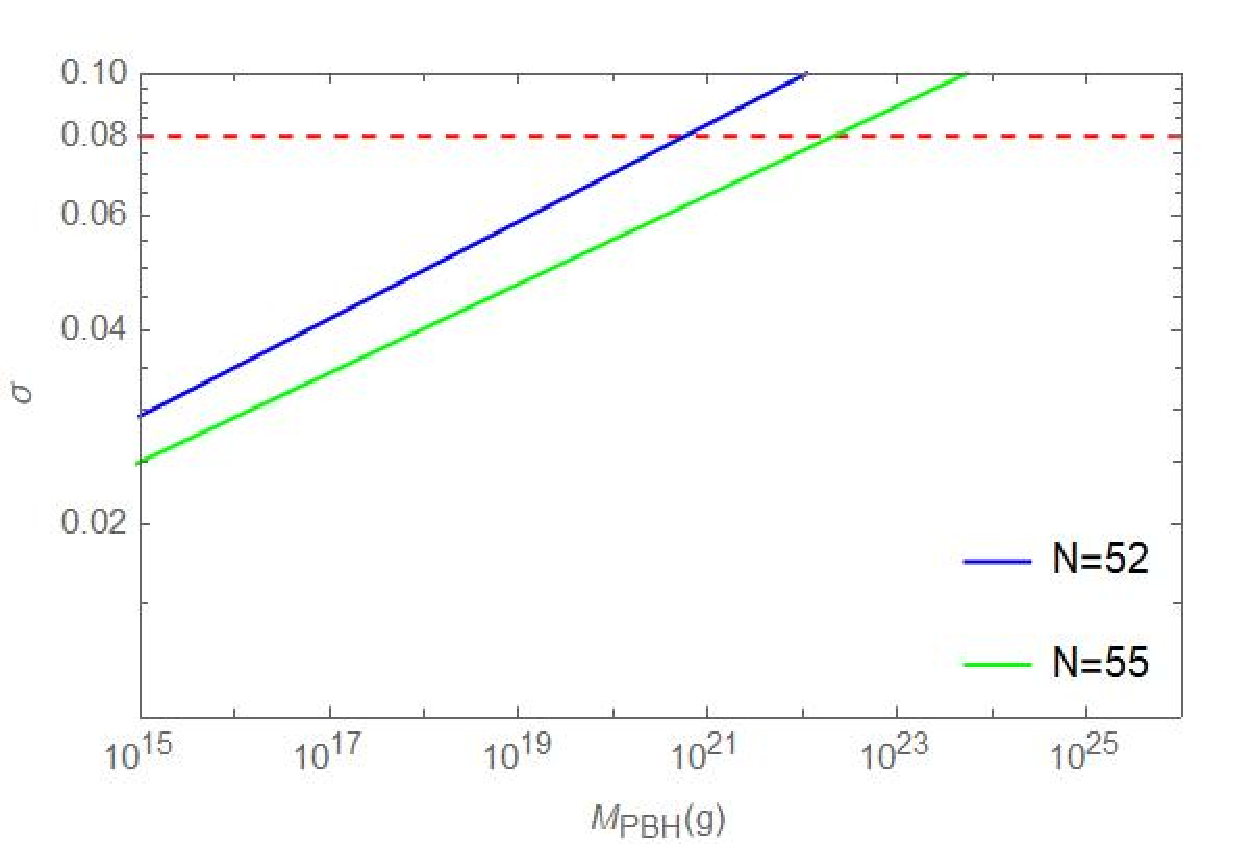}  
\caption*{-4.a-}
\end{subfigure}
\begin{subfigure}[b]{0.49\textwidth}
\includegraphics[width=1\linewidth]{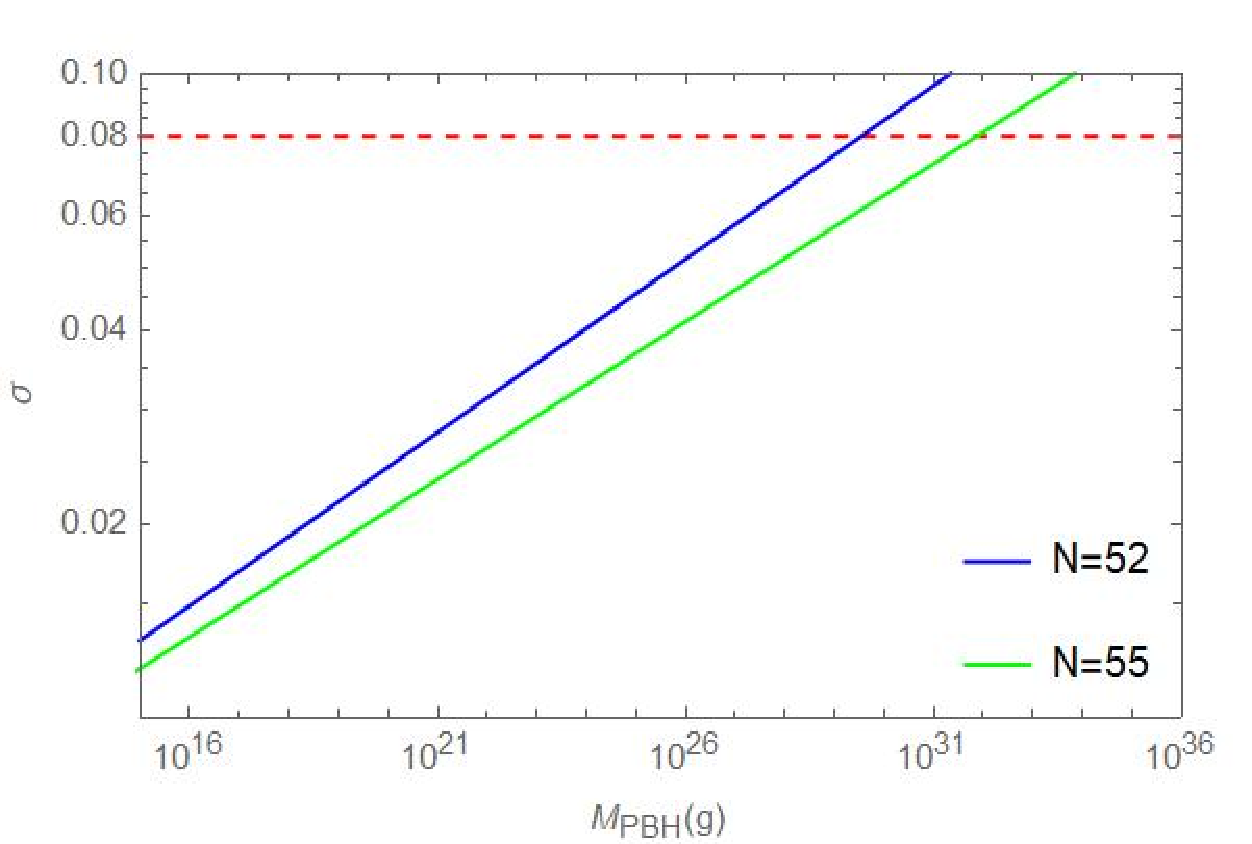} 
\caption*{-4.b-}
\end{subfigure} 
\caption{The mass variance against the PBH mass, $M_{PBH}$, for $N=52$ and $N=55$ . Two values of the coupling constant are considered: $\xi=10^{-4}$ (left panel) and $\xi=10^{-4.1}$ (right panel). The dashed line represents the threshold $\sigma_{thresh}=0.08$ above which PBHs would be overproduced.} 
\label{Fig4}
\end{figure}

 In Fig. \ref{Fig4}, we show the evolution of the mass variance as a function of the primordial black hole mass, $M_{PBH}$, for two specific values of the e-folding number, $N$, namely $52$ and $55$. We consider two cases for the coupling constant, $\xi$, specifically $\xi=10^{-4}$ (left panel) and $\xi=10^{-4.1}$ (right panel). The dashed line indicates the threshold $\sigma_{thresh}=0.08$ above which PBHs would be overproduced. For both scenarios, the mass variance, stays less than $\sigma_{thresh}$ for a mass range that satisfies the observational constraints. Additionally, for $\xi=10^{-4.1}$, the probability of generating more massive PBHs is higher compared to the case with $\xi=10^{-4}$. The influence of the number of e-folds is evident, as a simple increase in $N$ results in the production of more massive primordial black holes.     
\begin{figure}[h!]
\centering
\begin{subfigure}[b]{0.49\textwidth}
\includegraphics[width=1\linewidth]{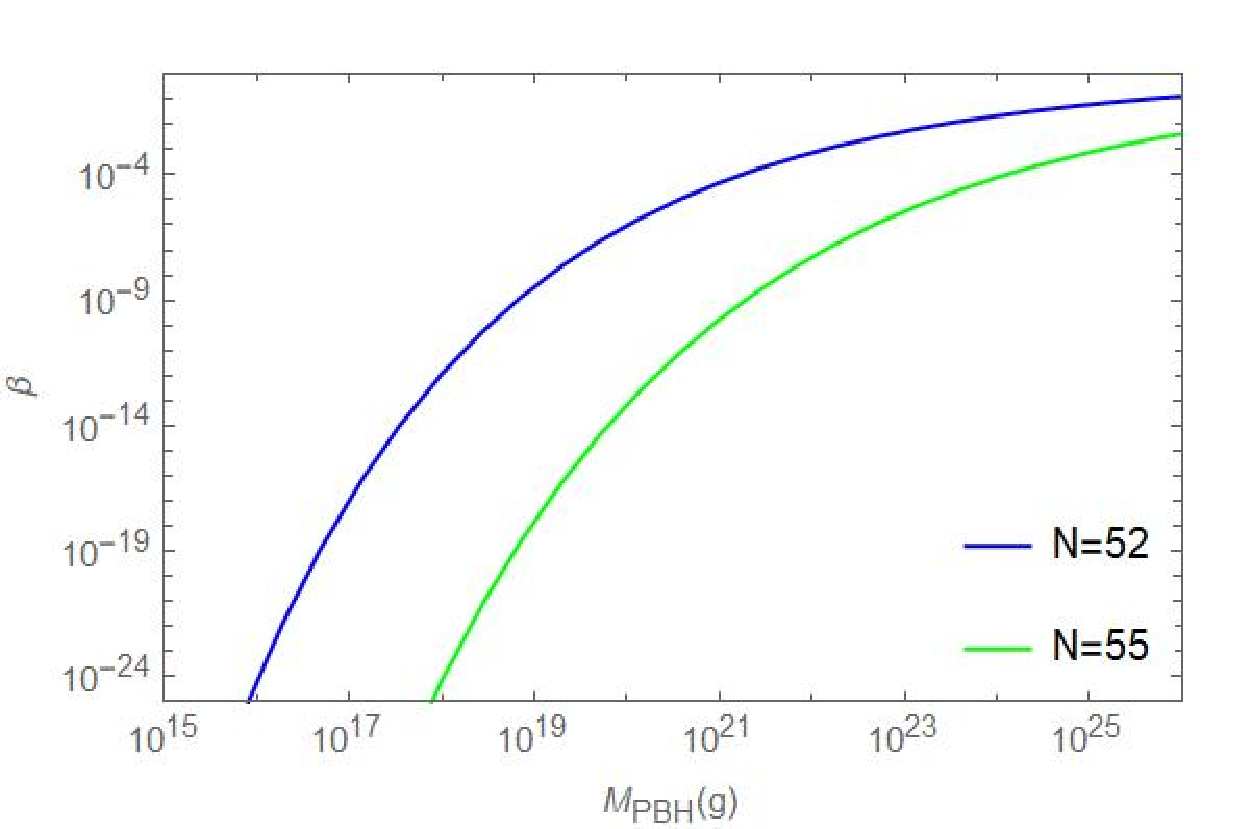} 
\caption*{-5.a-}
\end{subfigure}
\begin{subfigure}[b]{0.49\textwidth}
\includegraphics[width=1\linewidth]{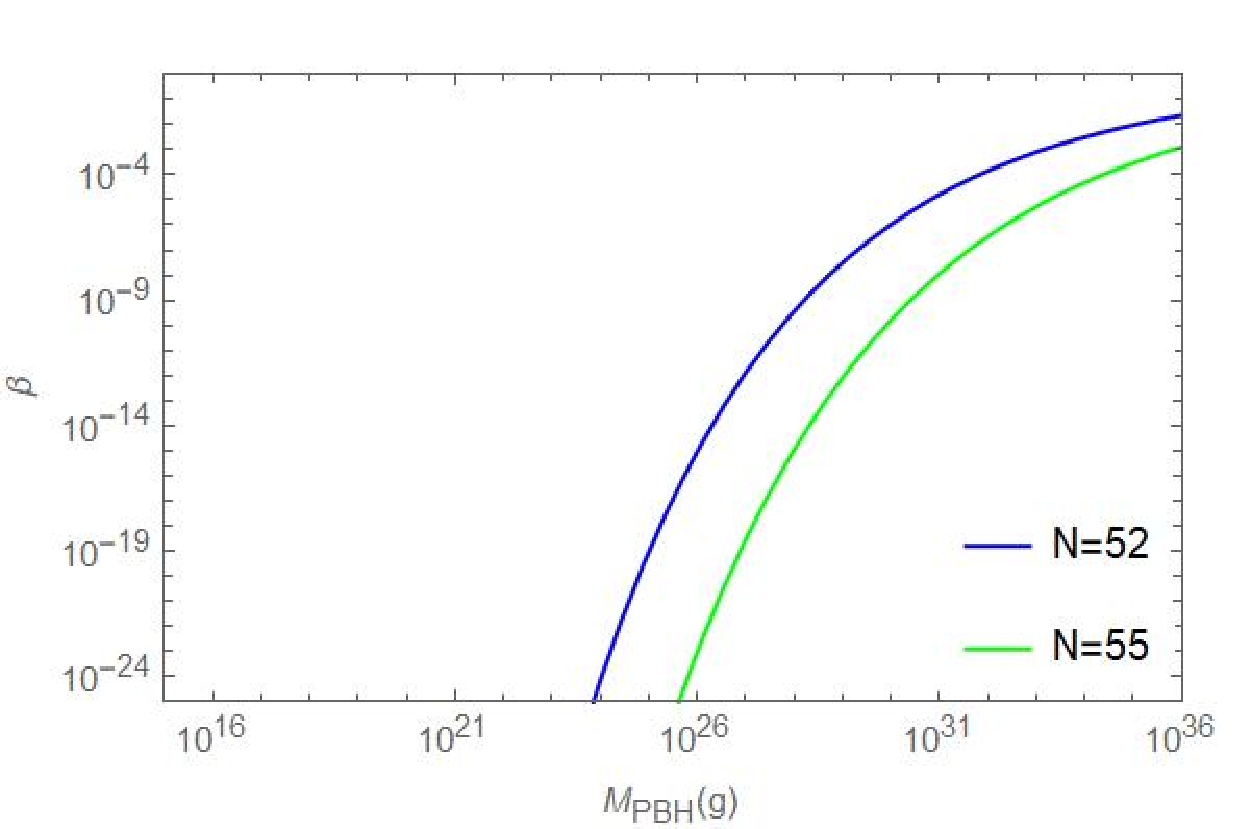} 
\caption*{-5.b-}
\end{subfigure} 
\caption{The fraction of the total energy density collapsing into PBHs as a function of the PBH mass, considering two cases where $N=52, 55$. We consider two coupling constant values, specifically $\xi=10^{-4}$ (left panel) and $\xi=10^{-4.1}$ (right panel).}
\label{Fig5}
\end{figure}
\par Fig. \ref{Fig5} illustrates the variation of the mass fraction, $\beta (M_{PBH})$, which represents the portion of the total energy density collapsing into primordial black holes, as a function of the PBH mass. Two different values of e-folding number are considered: $N=52$ (blue curve) and $N=55$ (green curve). The left and the right panels correspond to two different values of the coupling constant, $\xi=10^{-4}$ and $\xi=10^{-4.1}$, respectively. In both cases, the probability of PBHs formation is consistent with observational constraints, as the mass fraction, $\beta$, remains below the observational threshold of $10^{-20}$ within the allowed PBH mass range. Furthermore, all curves reveal that as the PBH mass increases, the function $\beta$ increases as well. It is also evident that an increase in the number of e-folds leads to the generation of more massive PBHs.
\section{Primordial black holes as a DM candidate}\label{secVI}
Primordial black holes satisfy all the necessary conditions to be considered as a viable DM candidate. They are cold, non-baryonic objects formed before the Big Bang nucleosynthesis, and stable over cosmological timescales. Furthermore, they can be produced in sufficient abundance to account for the observed dark matter. Although PBHs evaporate by emitting Hawking radiation \cite{Hawking:1975vcx}, they are considered stable if its initial mass is greater than $10^{15}$g \cite{Page:1976df}. Moreover, From the mass fraction function, $\beta$, one can compute the contribution of PBHs to the current total dark matter density. Returning to Eq.\eqref{eq3.15} and replacing $\rho_{tot}$ by its expression, we write 
\begin{eqnarray}
\beta\left(M_{PBH}\right)=\frac{\rho_{PBH}}{3 M^2_{pl} H^2}=\frac{\rho_{PBH}}{3 M^2_{pl} H^2_0}\left( \frac{H_0}{H}\right)^2.
\end{eqnarray}
Since primordial black holes behave like non-relativistic matter, PBH energy density is expressed as $\rho_{PBH}\propto a^{-3}$, implying that $\rho_{PBH}=\rho_{PBH,0}\left( a_0/a\right)^3$. Then, we write 
\begin{eqnarray}
\nonumber \beta \left(M_{PBH}\right)&=&\frac{\rho_{PBH,0}}{\rho_{DM,0}} \frac{\rho_{DM,0}}{3 M^2_{pl} H^2_0}\left( \frac{H_0}{H}\right)^2\left(\frac{a_0}{a}\right)^3\\
&=& f_{PBH} \Omega_{DM,0} \left( \frac{H_0}{H}\right)^2 a^{-3},
\end{eqnarray}
where $f_{PBH}$ and $\Omega_{DM,0}$ are the mass function describing fractional abundance of PBHs and the current density parameter of dark matter defined respectively as
\begin{eqnarray}
f_{PBH}\left( M_{PBH}\right)=\frac{\Omega_{PBH,0}}{\Omega_{DM,0}},
\end{eqnarray}
and 
\begin{eqnarray}
\Omega_{DM,0}=\frac{\rho_{DM,0}}{3 M^2_{pl} H^2_0}.
\end{eqnarray}
Substituting the expressions of $H$ and $a$ in terms of $\frac{M_{PBH}}{M_s}$ from equations \eqref{eq3.10} and \eqref{eq3.11}, one obtains \cite{Sasaki:2018dmp}
\begin{eqnarray}
f_{PBH}\left( M_{PBH}\right)=1.52\times 10^{8}\beta\left( M_{PBH}\right)\left(\frac{\gamma}{0.2}\right)^{1/2}\left(\frac{106.75}{g_{*}}\right)^{1/4}\left( \frac{M_{PBH}}{M_s}\right)^{-1/2},
\end{eqnarray} 
\nolinenumbers
\begin{figure}[h!]
\centering
\begin{subfigure}[b]{0.49\textwidth}
\includegraphics[width=1\linewidth]{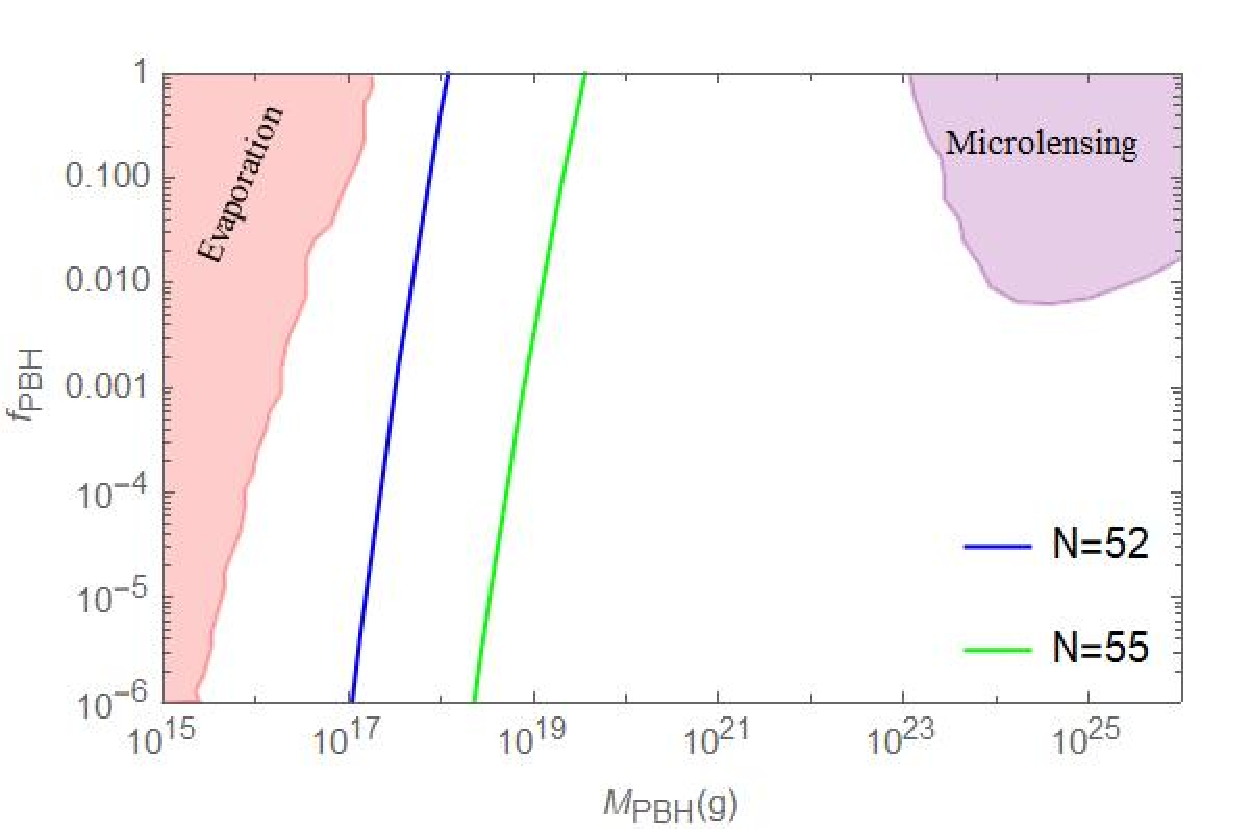}  
\caption*{-6.a-}
\end{subfigure}
\begin{subfigure}[b]{0.49\textwidth}
\includegraphics[width=1\linewidth]{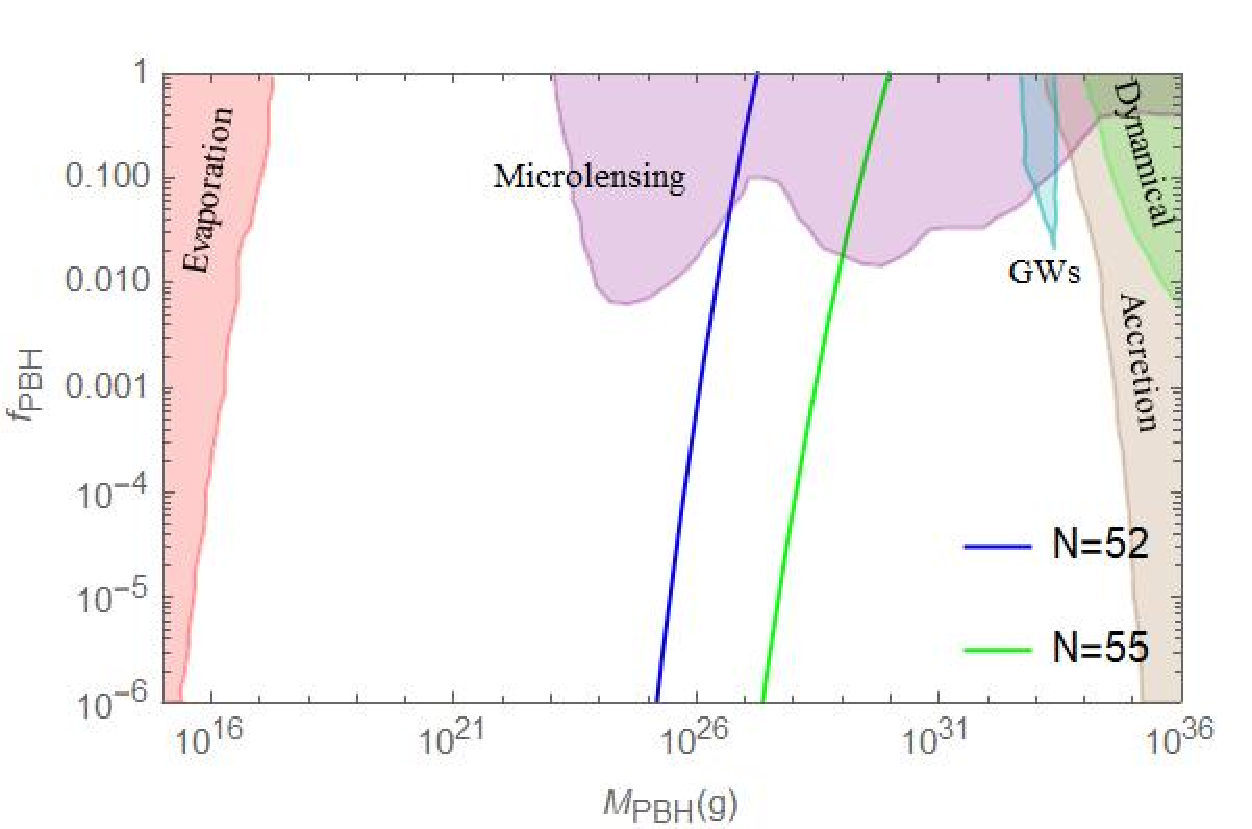}  
\caption*{-6.b-}
\end{subfigure} 
\caption{ The abundance of primordial black holes, $f_{PBH}$, as a function of the PBHs mass, for $N=52$ (blue curve) and $N=55$ (green curve), considering two cases where $\xi=10^{-4}$ (left panel) and $\xi=10^{-4.1}$ (right panel). The bounds depicted corresponding to observational constraints on PBHs abundance are evaporation (red region), microlensing (purple region), gravitational waves (cyan region) \cite{Raidal:2017mfl}, accretion (brown region) \cite{Serpico:2020ehh} and dynamical effects (green region) \cite{Inoue:2017csr}.}
\label{Fig6}
\end{figure}
where we have taken $\Omega_{DM,0}=0.26$. The abundance of primordial black holes as a function of the PBHs mass is shown in Fig. \ref{Fig6}, considering two values of the coupling constant, which are $\xi= 10^{-4}$ (left panel) and $\xi= 10^{-4.1}$ (right panel). We adopt two values of the e-folds number, $N$, specifically $N=52$ and $N=55$. In addition, we include the observational constraints on PBHs abundance from evaporation (red region), lensing (purple region), gravitational waves (cyan region) \cite{Raidal:2017mfl}, CMB distortions (brown region) \cite{Serpico:2020ehh}, and the accretion from X-ray binaries (green region) \cite{Inoue:2017csr} (see \cite{Tada:2023pue} for more details). For the first case, where $\xi= 10^{-4}$, both lines corresponding to $N=52$ and $N=55$ fall within allowed mass range, in which PBHs can account for 100\% of the present dark matter, between $1\times10^{17}$ and $3.75\times10^{19}$g. In contrast, for $\xi= 10^{-4.1}$, primordial black holes serve only as a fraction to the current dark matter abundance. Specifically, PBHs contribute about 4.5\% and 1.8\% for $N=52$ and $N=55$, respectively. In this case, the PBH masses span the range $1.36\times10^{25}\leq M_{PBH}\leq 1\times 10^{29}$g. As a key result, the PBH mass is highly sensitive to both the coupling constant, $\xi$ and the e-folding number $N$.
\section{Conclusion} \label{secV}
\par  In this paper, we investigated the possibility of primordial black holes generation within the framework of the Higgs hybrid metric-Palatini model, based on the non-minimal coupling between the inflaton field and the Palatini curvature. First, we have briefly reviewed the dynamics of our Higgs inflationary model. In particular, we focused on the key inflationary parameters such as the spectral index, $n_s$ and the tensor-to-scalar ratio, $r$. Our results show that the predicted evolution of these parameters is consistent with the latest observational constraints arising from the ACT, DESI, and Planck collaborations.
\par Additionally, we have studied the power spectrum of curvature perturbations. The behavior of curvature power spectrum exhibits a significant enhancement at small scales, corresponding to large wavenumbers $k$. As a result, gravitational collapse of over-dense regions can occur after inflation, leading to the production of PBHs during the radiation-dominated era. 
\par Furthermore, we investigated the probability of primordial black holes production and their contribution to the total energy density of the Universe. To this end, we analyzed the mass variance $\sigma(M_{PBH})$ and the mass fraction, $\beta(M_{PBH})$, considering two different values of both the coupling constant, $\xi$, and the e-folds number, $N$. We found that the probability of PBHs generation is consistent with constraints provided by observational data.
\par Finally, we studied the PBHs as potential candidates for the dark matter. For this purpose, we analyzed the abundance function $f(M_{PBH})$ considering two cases corresponding to two different values of the coupling constant, $\xi$, for two distinct values of the e-folds number, $N$. Our findings reveal that it is possible to generate primordial black holes in the allowed mass range that aligns with observations. Consequently, for $\xi=10^{-4}$ and $\xi=10^{-4.1}$, with $N=52$ and $N=55$, PBHs could account for either the entire dark matter content of the Universe or only a fraction of it, respectively.

\end{document}